\documentclass[referee]{raa}           

\usepackage{graphicx,times}
\usepackage{natbib}
\usepackage{amssymb,amsmath}
\bibpunct{(}{)}{;}{a}{}{,}
\usepackage[pagebackref=true]{hyperref}

\begin{document}

   \title{UV flux variation study in contact binary VW Cephei}

   \volnopage{ {\bf 20XX} Vol.\ {\bf X} No. {\bf XX}, 000--000}
   \setcounter{page}{1}

\author{Anurag Baruah
      \inst{*}, Mayukh Pahari\inst{}
   }
   \institute{Department of Physics, Indian Institute of Technology Hyderabad, Kandi, Sangareddy 502284, India; {\it ph24resch11002@iith.ac.in}\\
   \vs \no
   {\small Received 20XX Month Day; accepted 20XX Month Day}
}

\abstract{Despite many attempts, the origin of UV emission line and continuum in contact binary stars remains unclear. We present a substantial UV spectroscopic analysis of VW Cephei, a late-type contact binary system, using 46 low-resolution spectra from the International Ultraviolet Explorer (IUE) in the wavelength range 1150–1978~\AA. By modelling continuum and emissions lines in individual spectra, we report the significant detection of O\textsc{iii}] (1660 and 1666~\AA) and Si\textsc{iv} (1393 and 1402~\AA) line complexes. We observe that UV fluxes for both continuum and emission lines like C\textsc{iv}, O\textsc{iii}], C\textsc{ii} and Si\textsc{iv} vary significantly (fractional rms variability up to 45\%) from hours to years. In addition, line widths also change by hundreds of kilometres/sec. The UV flux variabilities observed in the continuum bands and line emissions are uncorrelated. 
However, most of the flux values follow the binary orbital period observed from optical data. Our analysis indicates that, while the variation in continuum flux may be attributed to a heated photosphere, the line width measurements indicate that the emission lines are likely formed in the dynamical clouds associated with Roche lobe overflow. We estimate the mass transfer rate of \(\dot{M} = (0.82 \pm 0.01) \times 10^{-7} \, M_{\odot}\,\mathrm{yr^{-1}}\) from UV line fluxes, which is in good agreement with optical studies.
}
   \maketitle
\section{Introduction}           
\label{sect:intro}
Contact binaries consist of two main-sequence stars rotating around a common centre of mass, in which both components fill their Roche lobes \citep{1959cbs..book.....K}. In such systems, energy is transferred from one star to the other through the inner Lagrangian point (L1), causing the two components to share a common envelope (CE) and to have nearly equal temperatures even though their masses are quite different \citep{1968ApJ...153..877L,1968ApJ...151.1123L}. They show continuous flux variation in their light curve. The orbital periods of most of these binaries are shorter than one day, and they follow the famous period-colour relation \citep{1967MmRAS..70..111E, 1998AJ....116.2998R}.

Contact binaries are divided into two main subtypes: the A-type (or early type), consisting of A or F stars and the W-type (or late type), made of G or K stars. The standard definition says that for A-type binaries, the primary minimum of eclipse is a transit. The more massive primary component has a higher surface brightness than the secondary one. In the case of W-type systems, the primary minimum corresponds to an occultation. Therefore, the secondary component has a higher mass than the primary, even though the primary has a higher temperature. These distinct physical processes in contact binaries make them a fascinating category of binary systems; studying these systems can enhance our understanding of binary star evolution and potentially contribute to broader theories of stellar evolution.

Many contact binaries show cyclic modulation, which may be due to a distant third component. \citet{1992ApJ...385..621A} proposed a mechanism that could be responsible for such cyclic period variations. According to his model, magnetic activity and cyclic variation of internal angular momentum change the gravitational quadrupole moment of the components, which in turn change the orbital period. Some contact binaries exhibit asymmetry in their light curve maxima, known as the O’Connell effect \citep{1951PRCO....2...85O, 1968AJS....73R..26M}, which is likely caused by starspots resulting from magnetic activity on the stellar surfaces. The variation between the maxima can fluctuate from one orbit to the next due to the dynamic nature and evolving distribution of these cooler active regions.

VW Cephei, one of the most frequently observed contact binaries, was discovered by \citet{1926ApJ....64..215S}. Its brightness (V = 7.30–7.84 mag) and short orbital period ($\sim$ 6.67 hours) make it an attractive target. The system is notable for exhibiting the O’Connell effect and for hosting a third component confirmed through astrometry \citep{1975AJ.....80..662H}. Despite extensive study, no model has yet fully accounted for all its observed properties. The unusual light curve behaviour of VW Cephei was first detected during dedicated photometric monitoring by \citet{1966BAN....18..448K}. Over the years, various models have been proposed to explain its asymmetric maxima, including a circumstellar ring \citep{1966BAN....18..448K}, a hot spot formed by gas streams \citep{1973A&A....26..357V, 1976AcA....26..319P}, precession of the stars' rotational axes \citep{1983A&A...128..391W}, and cool starspots \citep{1982Ap&SS..85...43Y}. Among these, the cool starspot model has been the most consistent with observations and widely adopted in light curve modeling. Some basic information of VW Cephei is presented in Table \ref{tab1}.
\begin{table}[htbp]
\centering
\caption{Basic Information of VW Cephei. Values of RA and Dec are taken from \citep{2021A&A...649A...1G} and all the other parameter values are from \citet{2018A&A...612A..91M}. \label{tab1}}
\setlength{\tabcolsep}{6pt}
\small
\begin{tabular}{ll}
\hline
Parameter & Value \\
\hline          
RA & 20h 37m 21.63s \\
Dec & +75d 36m 01.89s \\
Spectral type &	G8V+K0V\\
Mass-ratio & 0.302 $\pm$ 0.007\\
Inclination ($^{\circ}$) & 62.86 $\pm$ 0.04\\
Separation (a) [$10^6$km] & 1.412 $\pm$ 0.01\\
\hline
\end{tabular}
\end{table}
Through optical spectroscopic analysis of VW Cephei, \citet{2000ApJ...531..467H}, determined a mass ratio of, q = 0.395 ± 0.016. Their observations also revealed that both stellar components were heavily spotted, with the more massive primary exhibiting an off-centered polar spot during the time of study. X-ray spectroscopic studies of VW Cephei by \citet{2004A&A...415.1113G} and \citet{2006ApJ...650.1119H} revealed significant stellar activity. \citet{2004A&A...415.1113G} found that the system possesses an extended corona enveloping both components and displays flaring behavior. \citet{2006ApJ...650.1119H} confirmed the presence of flares and observed that the corona is predominantly concentrated near the polar regions of the primary star. In the ultraviolet regime, \citet{2014NewA...29...47S} analyzed emission lines and detected both short- and long-term variability in their strength, which they attributed to chromospheric activity associated with the primary component. Using IUE/LWP data \citet{1986ApJ...311..937V} found the UV continuum flux of VW Cephei follows the trend of its optical data.
This paper presents a detailed UV spectroscopic analysis of VW Cephei at short wavelengths and examines its continuum and emission line flux variation using the IUE data. The Si\textsc{iv} and O\textsc{iii}] emission lines are detected and analyzed for the first time in this study. The observations and data reduction methods are presented in Section \ref{sect:Obs}. Section \ref{sect:Analysis} outlines the results obtained from the spectral analysis. In Section \ref{sect:discussion}, the activity of VW Cephei based on the results is discussed. Finally, concluding remarks about this work are presented in Section \ref{sect:conclusion}.
\section{Observations and Data Reduction}
\label{sect:Obs}
International Ultraviolet Explorer (IUE) have performed 46 low-resolution short-wavelength spectral observations between July 10, 1978 and May 30, 1992, in the wavelength range of 1150-1978~\AA. We follow the data analysis methodology outlined by \citet{Wanders1997}. Raw spectral images are processed using both the TOMSIPS \citep{Ayres1993} and NEWSIPS \citep{Nichols1993} pipelines. For this study, we utilise the NEWSIPS-reduced spectra, in accordance with the recommendation of \citet{Wanders1997}, as the {\textsc TOMSIPS} reduction is affected by nonlinearities in wavelength calibration caused by uncorrected long-term drifts. The NEWSIPS pipeline, by contrast, yields spectra that are consistent with HST observations and does not require further calibration. Notably, the NEWSIPS pipeline introduces a minor wavelength shift of about 1–2~\AA~due to pointing inaccuracies with the large aperture \citep{Wanders1997}. To address this, an offset correction is applied to align the prominent C\textsc{iv} emission line consistently across all spectra. CCM89 \citep{1989ApJ...345..245C} model is used for extinction corrections. Due to the very low interstellar reddening E(B$-$V) = 0.0025, corrections are not significant.
\begin{table}[htbp]
\centering
\caption{IUE observations of VW Cephei. \label{tab2}}
\setlength{\tabcolsep}{6pt}
\small
\begin{tabular}{ccc|ccc}
\hline
Data ID & Obs Start Time & Exp Time & Data ID & Obs Start Time & Exp Time\\
 & (YYYY-MM-DD HH:MM:SS) & (s) & & (YYYY-MM-DD HH:MM:SS) & (s)\\
\hline
SWP01958 & 1978-07-10 12:38:00 & 3600 & SWP36606 & 1989-07-03 07:20:08 & 4800\\
SWP02005 & 1978-07-14 08:27:00 & 9000 & SWP36607 & 1989-07-03 09:23:23 & 5400\\
SWP06534 & 1979-09-16 09:11:24 & 9000 & SWP36608 & 1989-07-03 12:36:18 & 6000\\
SWP06535 & 1979-09-16 14:39:03 & 2760 & SWP36609 & 1989-07-03 15:32:25 & 6600\\
SWP07866 & 1980-02-04 15:32:11 & 3600 & SWP42527 & 1991-09-21 07:54:56 & 5400\\
SWP07867 & 1980-02-04 17:05:47 & 3600 & SWP42528 & 1991-09-21 10:30:03 & 5400\\
SWP30714 & 1987-04-05 11:50:02 & 6900 & SWP42531 & 1991-09-22 01:47:07 & 5400\\
SWP30715 & 1987-04-05 14:59:53 & 5100 & SWP42532 & 1991-09-22 05:05:04 & 5400\\
SWP33921 & 1988-07-15 03:42:33 & 4800 & SWP42533 & 1991-09-22 07:30:06 & 5400\\
SWP33922 & 1988-07-15 05:48:35 & 4500 & SWP42534 & 1991-09-22 10:07:12 & 5400\\
SWP33923 & 1988-07-15 07:47:31 & 4800 & SWP42535 & 1991-09-22 12:36:43 & 5400\\
SWP33924 & 1988-07-15 11:22:58 & 4800 & SWP42544 & 1991-09-24 02:24:45 & 4200\\
SWP33926 & 1988-07-15 15:44:39 & 5340 & SWP42545 & 1991-09-24 05:36:37 & 5400\\
SWP36598 & 1989-07-02 06:13:53 & 3300 & SWP42546 & 1991-09-24 09:52:15 & 5400\\
SWP36599 & 1989-07-02 09:33:43 & 6600 & SWP44811 & 1992-05-30 07:33:25 & 4200\\
SWP36600 & 1989-07-02 12:22:53 & 6600 & SWP44812 & 1992-05-30 10:27:07 & 6480\\
SWP36601 & 1989-07-02 15:46:33 & 6600 & SWP44813 & 1992-05-30 13:23:43 & 5100\\
SWP36605 & 1989-07-03 04:47:01 & 5100 & & & \\
\hline
\end{tabular}
\end{table}

Table \ref{tab2} lists the IUE observations of VW Cephei. Each spectrum was carefully examined within the 1279–1797~\AA~wavelength range to screen for excessive noise, overexposure, or underexposure. Based on signal-to-noise considerations, only 35 of the 46 observations are found to be of sufficient quality for reliable spectral fitting.

\section{UV Spectral analysis of VW Cephei}
\label{sect:Analysis}
\begin{figure}[htbp]
    \centering
    \includegraphics[scale=0.7]{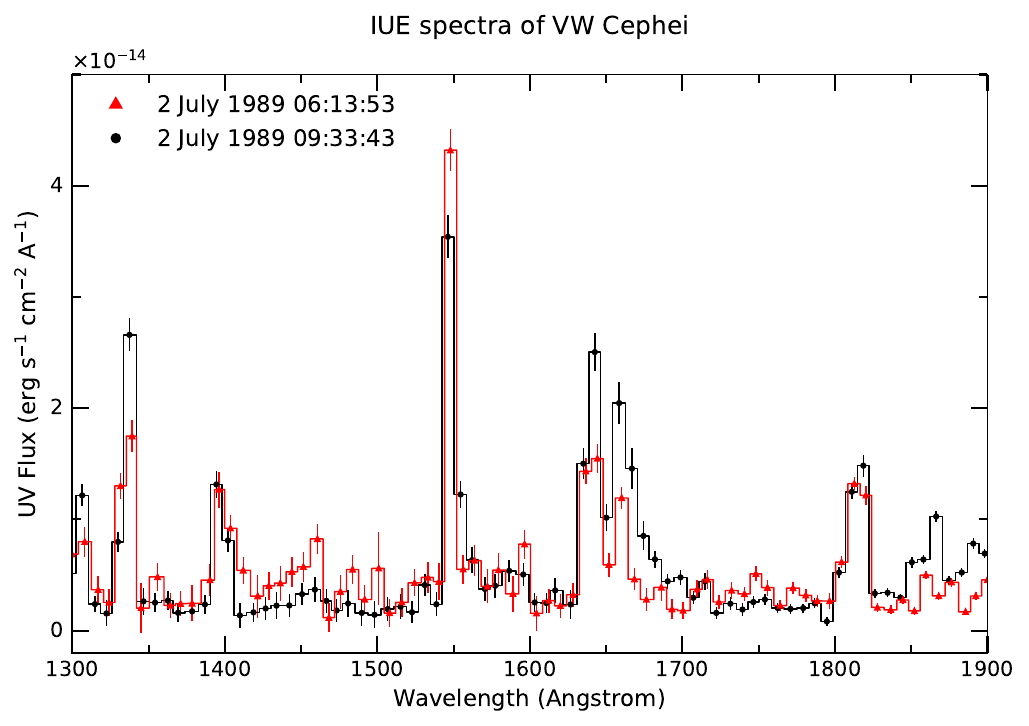}
    \caption{UV variability of VW Cephei as observed on 2 July 1989 at 06:13:53 (triangles) and 2 July 1989 at 09:33:43 (circles). For better visibility of continuum and line flux variations in short timescales, IUE spectra are plotted after binning to a coarse resolution (11.8~\AA~) in the wavelength range of 1300-1900~\AA.}
    \label{Fig1}
\end{figure}

There is significant variability in both line and continuum flux levels, along with line widths, for IUE ultraviolet spectra. We have presented such an example in Figure \ref{Fig1}, where the flux observed on 2 July 1989 at 06:13:53 is notably different from that recorded just three hours later.

\begin{figure*}[htbp]
    \centering
    \includegraphics[width=\textwidth]{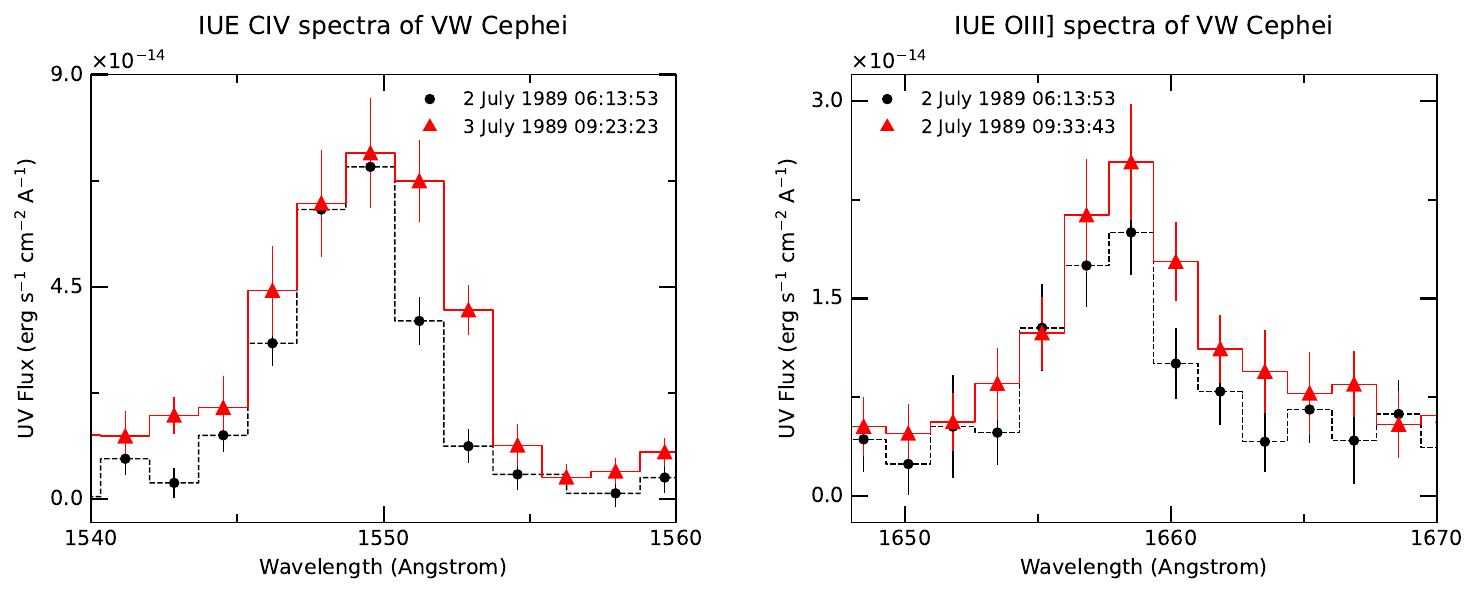}
    \caption{UV line width variability of VW Cephei. Left: Width of C\textsc{iv} line as observed on 3 July 1989 at 09:23:23 is broader than that of 2 July 1989 at 06:13:53. Right: Width of O\textsc{iii}] line as observed on 2 July 1989 at 09:33:43 is broader than that of 2 July 1989 at 06:13:53. \label{Fig5}}
\end{figure*}

Similarly, in Figure \ref{Fig5} we can see that the width of C\textsc{iv} as observed on 3 July 1989 at 09:23:23 is broader than that of 2 July 1989 at 06:13:53. Similarly, the width of O\textsc{iii}] line as observed on 2 July 1989 at 09:33:43 is broader than that of 2 July 1989 at 06:13:53.

\begin{figure}[htbp]
   \includegraphics[height=0.94\columnwidth, angle=270]{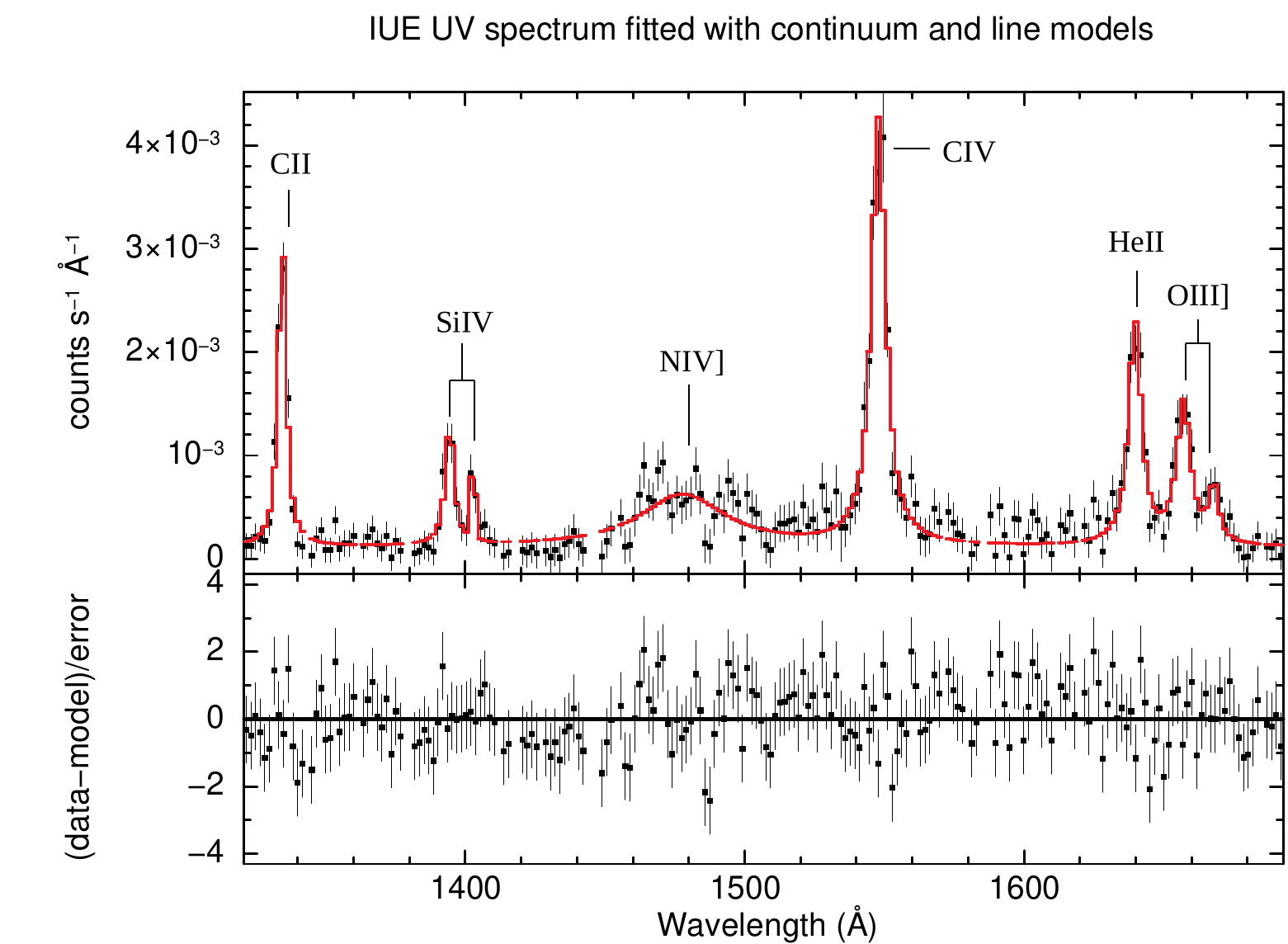}
   \caption{The best-fit ($\chi^2$/d.o.f. = 172.45/177) UV spectrum (1320-1690~\AA) is shown for the Obs ID SWP02005. The best-fit model (red) consists of continuum and emission line features, and the residual of the fitting is shown in the bottom panel. A similar fitting procedure is applied to all individual UV spectra.} 
   \label{Fig2}
\end{figure}
To extract the UV continuum and line light curves from the spectra, we employ a $\chi^2$ minimisation technique using \textsc{XSpec}. To enhance the signal-to-noise ratio, we bin the spectra by combining every two consecutive wavelength points. Each spectrum is fitted with a suitable combination of a power-law function—representing the underlying continuum—and multiple narrow and broad Lorentzian components to model the emission lines. The best-fitting model for the observation SPW02005 yields residuals with a reduced chi-square value of $\chi^2$/d.o.f. = 172.45/177. The time-averaged UV spectrum along with the model components and residuals is shown in Figure \ref{Fig2}, and the corresponding best-fit parameters are listed in Table \ref{tab3}. From the best-fitting model, the integrated fluxes of emission lines and the UV continuum are computed using the \textsc{cflux} convolution model in \textsc{XSpec}, which provides both the flux values over a specified wavelength range and their associated 1$\sigma$ uncertainties. We calculated the continuum flux in the range 1343-1390~\AA~(CONT1) and 1565-1633~\AA~(CONT2) and line fluxes at 1336 (C\textsc{ii}), 1393 (Si\textsc{iv}), 1550 (C\textsc{iv}) and 1660 (O\textsc{iii}])~\AA. Each of 35 spectra is fitted separately with the best-fitting model to obtain the continuum and line fluxes from each pointing.

\begin{figure*}[htbp]
    \includegraphics[width=\textwidth]{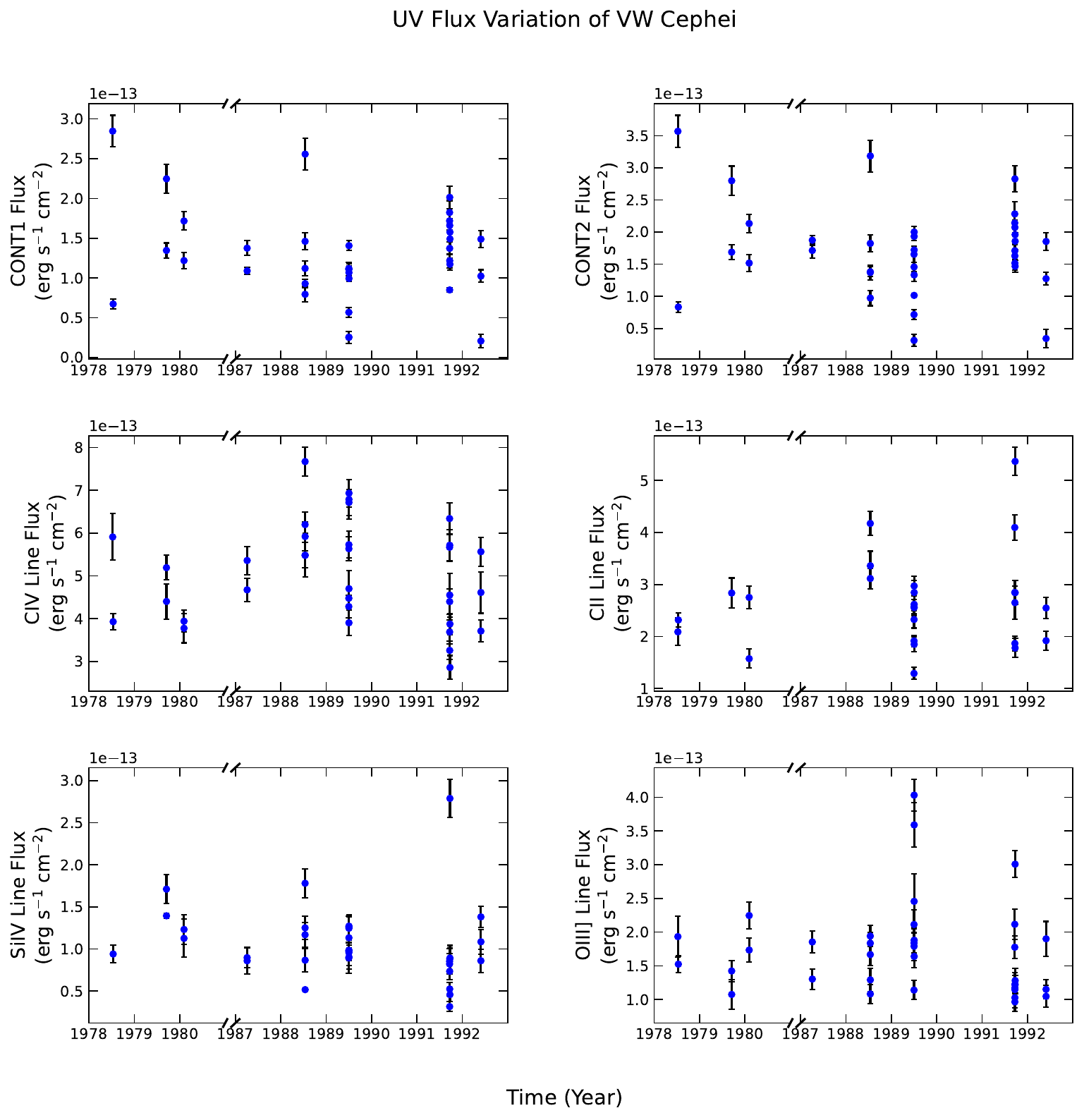}
    \caption{UV continuum, C\textsc{iv}, C\textsc{ii}, Si\textsc{iv} and O\textsc{iii}] line flux variation of IUE SWP data from 1978 to 1992. CONT1 and CONT2 represents the continuum flux in the range 1343-1390~\AA~and 1565-1633~\AA~respectively.}
    \label{Fig3}
\end{figure*}
\begin{figure}[htbp]
    \centering
    \includegraphics[width=\textwidth]{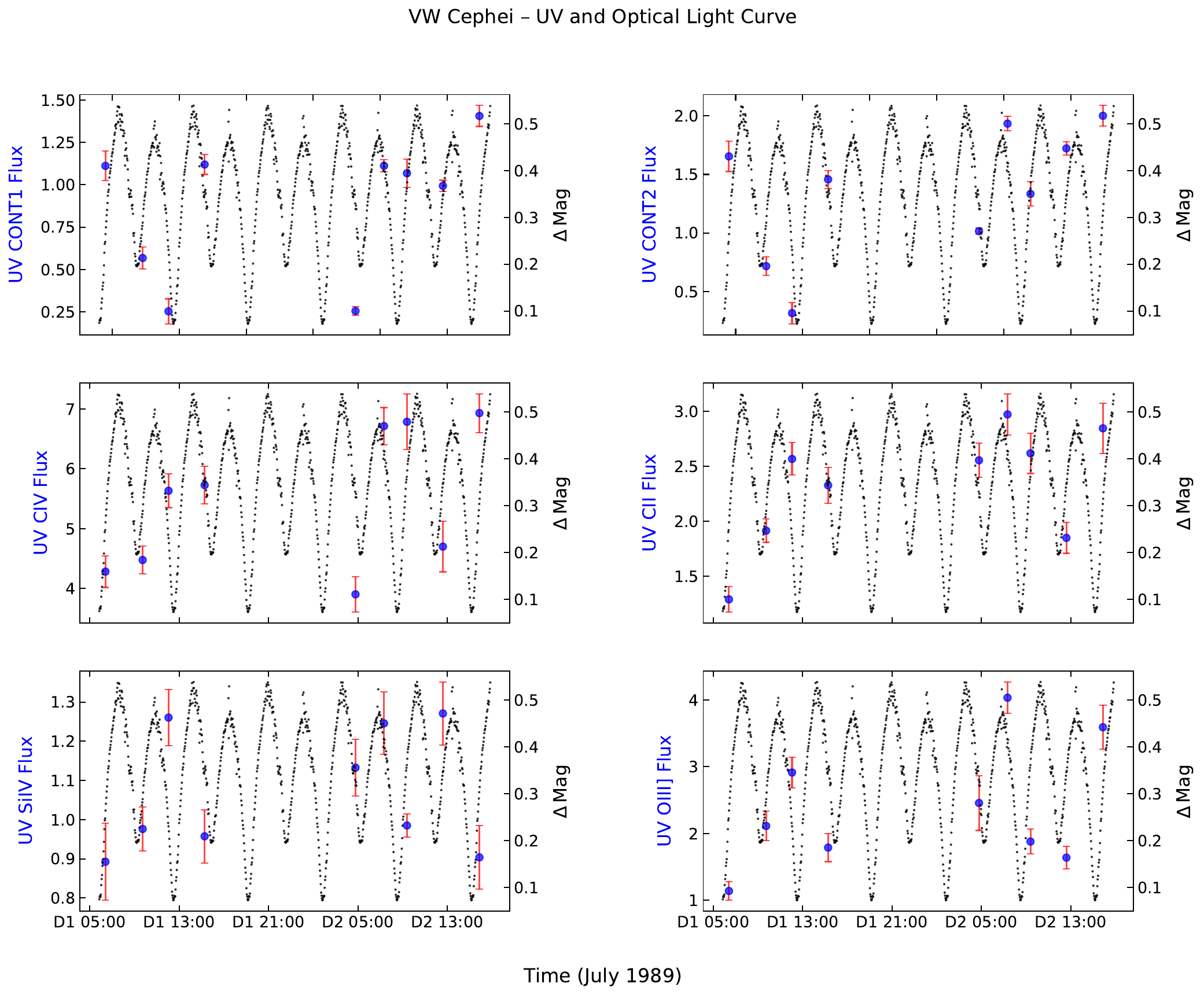}
    \caption{UV continuum and line flux variability of VW Cephei as observed on 1989. The flux axis has unit $10^{-13}\ \mathrm{erg\ s^{-1}\ cm^{-2}\ }$. UV flux (blue dot) plotted alongside optical data (black dot) in $\Delta$Mag scale. The optical data is taken from \citet{1981AcA....31..351K} and extrapolated to our required timescale using a period of 6.67 hours and an epoch of HJD 2444157.4131, corresponding to the primary minimum (E = 0) on 10 October 1979. D1 and D2 correspond to day 1 (2 July 1989) and day 2 (3 July 1989), respectively.}
    \label{Fig4}
\end{figure}

Figure \ref{Fig3} presents the variation of UV continuum and emission line flux with time, revealing significant variability in both components. Corresponding values are given in Table \ref{tab4}. For better visibility and to enable comparison with the optical light curve, we have shown the UV flux for observations of the year 1989 alongside the corresponding optical magnitudes from the same period in Figure \ref{Fig4}. The optical data is taken from \citet{1981AcA....31..351K} and extrapolated to our required timescale using a period of 6.67 hours.
\begin{table}[htbp]
\centering
\caption{Best fit parameters for the IUE spectra of VW Cephei on 14 July 1978 at 08:27:00 (SWP02005).\label{tab3}}
\setlength{\tabcolsep}{6pt}
\small
\begin{tabular}{ccll}
\hline
Emission Line & Component & Parameter & Value\\
\hline
C\textsc{iv}  & lorentz  & LineE ($10^{-3}$keV) & $8.01\pm0.01$\\
      &          & Width ($10^{-5}$keV) & $3.05^{+0.28}_{-0.19}$ \\
      &          & norm                 & $0.04\pm0.02$ \\
C\textsc{ii}  & lorentz  & LineE ($10^{-3}$keV) & 9.29 ± 0.01 \\
      &          & Width ($10^{-5}$keV) & $2.14^{+0.15}_{-0.23}$\\
      &          & norm                 & 0.02 ± 0.01 \\
Si\textsc{iv}$^1$ & lorentz  & LineE ($10^{-3}$keV) & $8.89 \pm 0.03$ \\
      &          & Width ($10^{-5}$keV) & $2.15^{+0.45}_{-0.33}$ \\
      &          & norm                 & 0.01 ± 0.01 \\
Si\textsc{iv}$^2$ & lorentz  & LineE ($10^{-3}$keV) & $8.83^{+0.06}_{-0.03}$\\
      &          & Width ($10^{-4}$keV) & $2.25^{+0.08}_{-0.02}$\\
      &          & norm                 & 0.01 ± 0.01\\
He\textsc{ii} & lorentz  & LineE ($10^{-3}$keV) & 7.56 ± 0.01 \\
      &          & Width ($10^{-5}$keV) & $2.39^{+0.35}_{-0.14}$\\
      &          & norm                 & 0.02 ± 0.01\\
O\textsc{iii}]$^1$& lorentz  & LineE ($10^{-3}$keV) & $7.48^{+0.13}_{-0.19}$\\
      &          & Width ($10^{-5}$keV) & $2.61\pm0.28$\\
      &          & norm                 & 0.01 ± 0.01\\
O\textsc{iii}]$^2$& lorentz  & LineE ($10^{-3}$keV) & $7.43\pm0.03$\\
      &          & Width ($10^{-5}$keV) & $1.67^{+0.74}_{-0.41}$\\
      &          & norm                 & 0.03 ± 0.01\\
N\textsc{iv}] & lorentz  & LineE ($10^{-3}$keV) & $8.36\pm0.01$\\
      &          & Width ($10^{-4}$keV) & $2.13\pm0.25$\\
      &          & norm                 & 0.03 ± 0.02\\
      & powerlaw & PhoIndex             & 1.93 ± 0.03 \\
          &          & norm             & $0.02\pm0.01$ \\
\hline
\end{tabular}
\end{table}
In Table~\ref{tab:SD}, we present the average flux, standard deviation (SD), root mean square (RMS) and fractional RMS for each continuum and emission line light curve. The RMS and fractional RMS are calculated using the standard formulas as defined in the following equations, where $x_i$ denotes the individual flux measurements and $\sigma_i$ represents the corresponding flux uncertainties. N is the total number of flux data points.
\[
    \mathrm{RMS} = \sqrt{\frac{1}{N} \sum_{i=1}^{N} x_i^2},\ \ \ \ \ \ \
     F_{\mathrm{rms}} = \frac{\sqrt{SD^2 - \langle \sigma_{\mathrm{i}}^2 \rangle}}{\langle x \rangle}.
\]

The relatively large SD values across all light curves indicate substantial variability in both the UV continuum and emission lines. Furthermore, high fractional RMS values, signify that the amplitude of variability is comparable to the average flux level, reflecting strong intrinsic fluctuations in the source.

\begin{figure*}
    \centering
    \includegraphics[scale=0.7]{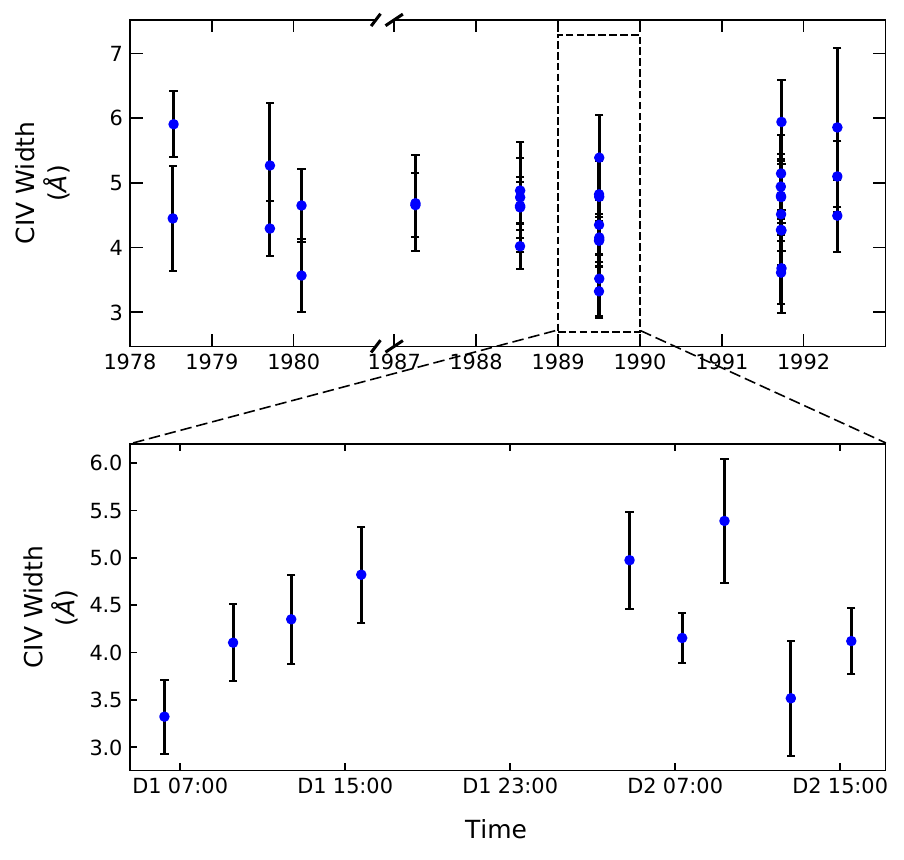}
    \caption{Variation of C\textsc{iv} line width over time. Top: Long-term variation (over years). Bottom: Short-term variation. D1 and D2 correspond to day 1 (2 July 1989) and day 2 (3 July 1989), respectively. \label{Fig6}}
\end{figure*}

\begin{figure*}[htbp]
    \centering
    \includegraphics[scale=0.7]{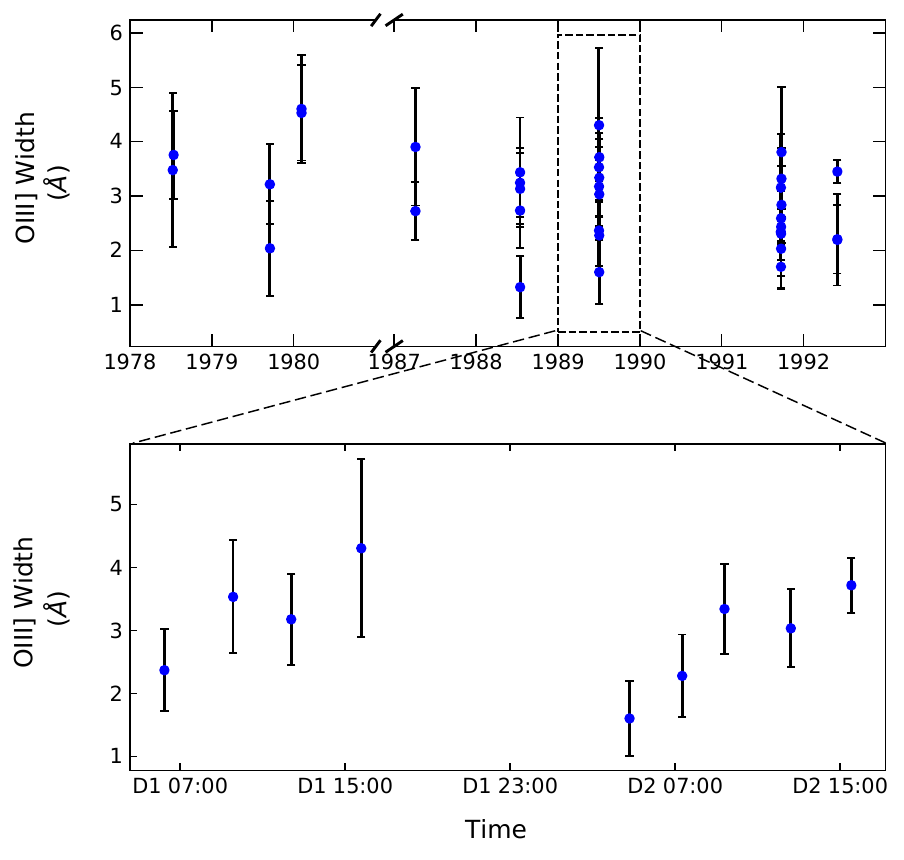}
    \caption{Variation of O\textsc{iii}] line width over time. Top: Long-term variation (over years). Bottom: Short-term variation. D1 and D2 correspond to day 1 (2 July 1989) and day 2 (3 July 1989), respectively. \label{Fig7}}
\end{figure*}

\clearpage
\begin{table*}[htbp]
\centering
\caption{Fluxes from SWP spectral modelling in $10^{-13}\ \mathrm{erg\ s^{-1}\ cm^{-2}\ }$ along with respective reduced chi-square values. Continuum fluxes are calculated within the wavelength range of 1343-1390~\AA~(CONT1) and 1565-1633~\AA~(CONT2). ND: Not Detected. \label{tab4}}
\begin{tabular}{cccccccc}
\hline
Data ID  & CONT1 flux & CONT2 flux & C\textsc{iv} flux & C\textsc{ii} flux & Si\textsc{iv} flux & O\textsc{iii}] flux & $\chi^2$/dof\\
\hline
SWP01958 & $2.85 \pm 0.20$ & $3.57 \pm 0.25$ & $5.91 \pm 0.54$ & $2.09 \pm 0.26$ & ND & $1.93 \pm 0.30$ & $120.77/163$\\
SWP02005 & $0.67 \pm 0.07$ & $0.83 \pm 0.08$ & $3.93 \pm 0.19$ & $2.32 \pm 0.13$ & $0.94 \pm 0.11$ & $1.52 \pm 0.12$ & $172.45/177$\\
SWP06534 & $1.35 \pm 0.09$ & $1.69 \pm 0.12$ & $5.19 \pm 0.29$ & $0.02 \pm 0.01$ & $1.71 \pm 0.17$ & $1.42 \pm 0.16$ & $214.32/195$\\
SWP06535 & $2.25 \pm 0.18$ & $2.80 \pm 0.23$ & $4.40 \pm 0.42$ & $2.84 \pm 0.28$ & $1.39 \pm 0.03$ & $1.07 \pm 0.22$ & $148.34/149$\\
SWP07866 & $1.22 \pm 0.10$ & $1.52 \pm 0.13$ & $3.78 \pm 0.34$ & $1.57 \pm 0.18$ & $1.13 \pm 0.23$ & $1.73 \pm 0.18$ & $112.32/176$\\
SWP07867 & $1.72 \pm 0.12$ & $2.13 \pm 0.14$ & $3.94 \pm 0.25$ & $2.75 \pm 0.21$ & $1.23 \pm 0.17$ & $2.25 \pm 0.20$ & $174.44/183$\\
SWP30714 & $1.37 \pm 0.09$ & $1.71 \pm 0.12$ & $4.67 \pm 0.27$ & $0.02 \pm 0.01$ & $0.90 \pm 0.12$ & $1.85 \pm 0.16$ & $178.09/175$\\
SWP30715 & $1.09 \pm 0.04$ & $1.87 \pm 0.08$ & $5.36 \pm 0.33$ & $0.02 \pm 0.02$ & $0.86 \pm 0.16$& $1.30 \pm 0.15$ & $266.30/163$\\
SWP33921 & $1.12 \pm 0.09$ & $1.37 \pm 0.12$ & $6.20 \pm 0.29$ & $3.11 \pm 0.19$ & $1.17 \pm 0.14$ & $1.94 \pm 0.16$ & $179.07/164$\\
SWP33922 & $0.79 \pm 0.10$ & $0.97 \pm 0.12$ & $7.67 \pm 0.33$ & $4.17 \pm 0.23$ & $1.25 \pm 0.14$ & $1.84 \pm 0.16$ & $183.19/183$\\
SWP33923 & $1.46 \pm 0.11$ & $1.82 \pm 0.13$ & $5.48 \pm 0.29$ & $0.03 \pm 0.02$ & $1.78 \pm 0.17$ & $1.67 \pm 0.17$ & $172.61/169$\\
SWP33924 & $0.93 \pm 0.05$ & $1.38 \pm 0.08$ & $5.92 \pm 0.34$ & $0.02 \pm 0.01$ & $0.87 \pm 0.14$ & $1.08 \pm 0.14$ & $161.94/173$\\
SWP33926 & $2.56 \pm 0.20$ & $3.18 \pm 0.25$ & $5.48 \pm 0.51$ & $3.36 \pm 0.28$ & $0.52 \pm 0.02$ & $1.29 \pm 0.18$ & $122.67/159$\\
SWP36598 & $1.11 \pm 0.09$ & $1.65 \pm 0.13$ & $4.28 \pm 0.26$ & $1.29 \pm 0.16$ & $0.89 \pm 0.18$ & $1.14 \pm 0.14$ & $135.28/172$\\
SWP36599 & $0.57 \pm 0.06$ & $0.72 \pm 0.08$ & $4.48 \pm 0.22$ & $1.91 \pm 0.11$ & $0.98 \pm 0.10$ & $2.11 \pm 0.22$ & $212.35/177$\\
SWP36600 & $0.25 \pm 0.07$ & $0.32 \pm 0.09$ & $5.63 \pm 0.28$ & $2.57 \pm 0.15$ & $1.26 \pm 0.13$ & $2.92 \pm 0.14$ & $217.09/172$\\
SWP36601 & $1.12 \pm 0.06$ & $1.46 \pm 0.08$ & $5.73 \pm 0.31$ & $2.33 \pm 0.16$ & $0.96 \pm 0.16$ & $1.79 \pm 0.21$ & $232.11/146$\\
SWP36605 & $0.25 \pm 0.01$ & $1.02 \pm 0.03$ & $3.90 \pm 0.29$ & $2.55 \pm 0.16$ & $1.13 \pm 0.12$ & $2.46 \pm 0.41$ & $177.73/173$\\
SWP36606 & $1.11 \pm 0.04$ & $1.93 \pm 0.06$ & $6.71 \pm 0.31$ & $2.97 \pm 0.18$ & $1.25 \pm 0.14$ & $4.03 \pm 0.24$ & $279.14/175$\\
SWP36607 & $1.07 \pm 0.08$ & $1.33 \pm 0.10$ & $6.79 \pm 0.46$ & $2.62 \pm 0.18$ & $0.98 \pm 0.12$ & $1.88 \pm 0.19$ & $220.30/173$\\
SWP36608 & $0.99 \pm 0.03$ & $1.72 \pm 0.06$ & $4.70 \pm 0.42$ & $1.85 \pm 0.14$ & $1.27 \pm 0.14$ & $1.64 \pm 0.17$ & $262.57/173$\\
SWP36609 & $1.41 \pm 0.06$ & $2.00 \pm 0.09$ & $6.93 \pm 0.32$ & $2.85 \pm 0.23$ & $0.90 \pm 0.14$ & $3.59 \pm 0.33$ & $262.92/162$\\
SWP42527 & $1.72 \pm 0.12$ & $2.14 \pm 0.15$ & $5.67 \pm 0.31$ & $4.10 \pm 0.24$ & $0.82 \pm 0.12$ & $1.78 \pm 0.17$ & $214.44/167$\\
SWP42528 & $1.82 \pm 0.15$ & $2.28 \pm 0.19$ & $6.34 \pm 0.37$ & $2.85 \pm 0.22$ & $0.85 \pm 0.15$ & $2.12 \pm 0.22$ & $125.65/164$\\
SWP42531 & $0.85 \pm 0.03$ & $1.63 \pm 0.05$ & $4.40 \pm 0.29$ & $1.87 \pm 0.15$ & $0.31 \pm 0.06$ & $1.17 \pm 0.11$ & $167.56/169$\\
SWP42532 & $1.37 \pm 0.08$ & $1.71 \pm 0.10$ & $3.69 \pm 0.27$ & $0.03 \pm 0.02$ & $0.73 \pm 0.10$ & $1.15 \pm 0.12$ & $234.87/171$\\
SWP42533 & $1.22 \pm 0.09$ & $1.52 \pm 0.11$ & $3.26 \pm 0.22$ & $0.03 \pm 0.02$ & $0.52 \pm 0.08$ & $0.96 \pm 0.14$ & $207.21/164$\\
SWP42534 & $2.01 \pm 0.14$ & $2.83 \pm 0.20$ & $5.72 \pm 0.37$ & $2.65 \pm 0.32$ & ND & $1.22 \pm 0.17$ & $128.77/172$\\
SWP42535 & $1.66 \pm 0.09$ & $2.07 \pm 0.11$ & $4.55 \pm 0.51$ & $0.02 \pm 0.01$ & $0.88 \pm 0.13$ & $1.03 \pm 0.15$ & $230.96/183$\\
SWP42544 & $1.58 \pm 0.10$ & $1.96 \pm 0.13$ & $3.86 \pm 0.25$ & $5.37 \pm 0.27$ & $2.79 \pm 0.23$ & $3.01 \pm 0.20$ & $255.95/194$\\
SWP42545 & $1.17 \pm 0.08$ & $1.47 \pm 0.09$ & $3.88 \pm 0.25$ & $1.78 \pm 0.18$ & $0.46 \pm 0.08$ & $1.22 \pm 0.14$ & $150.14/188$\\
SWP42546 & $1.49 \pm 0.10$ & $1.86 \pm 0.13$ & $2.86 \pm 0.28$ & $2.84 \pm 0.23$ & $0.89 \pm 0.15$ & $1.28 \pm 0.18$ & $179.31/176$\\
SWP44811 & $0.21 \pm 0.08$ & $0.34 \pm 0.14$ & $5.56 \pm 0.33$ & $0.03 \pm 0.01$ & $0.86 \pm 0.14$ & $1.15 \pm 0.14$ & $182.59/170$\\
SWP44812 & $1.02 \pm 0.08$ & $1.28 \pm 0.10$ & $3.71 \pm 0.25$ & $2.55 \pm 0.20$ & $1.38 \pm 0.13$ & $1.05 \pm 0.16$ & $235.45/189$\\
SWP44813 & $1.49 \pm 0.11$ & $1.85 \pm 0.14$ & $4.61 \pm 0.48$ & $1.92 \pm 0.18$ & $1.08 \pm 0.15$ & $1.90 \pm 0.26$ & $162.17/168$\\
\hline
\end{tabular}
\end{table*}
Our analysis also shows that there is noteworthy variation in width of emission lines. This temporal variation of line width is calculated for C\textsc{iv}, C\textsc{ii}, Si\textsc{iv} and O\textsc{iii}] lines and plotted for C\textsc{iv} and O\textsc{iii}] lines in Figure \ref{Fig6} and \ref{Fig7} respectively. Corresponding values of velocity widths due to Doppler broadening, for 2-3 July, 1989, are calculated using the relation \(\Delta v=\frac{\Delta \lambda}{\lambda}.c\) and presented in Table \ref{tab5}.

\begin{table}
\centering
\caption{Mean flux, standard deviation (SD), root mean square (RMS) and fractional RMS for the continuum and emission line light curves. The flux, SD, and RMS values are expressed in units of $10^{-13}\ \mathrm{erg\ s^{-1}\ cm^{-2}}$. \label{tab:SD}}
\begin{tabular}{ccccc}
\hline
Lightcurve type & Mean Flux & Standard Deviation & RMS & Fractional RMS\\
\hline
CONT1 & $1.28 \pm 0.02$ & $0.59 \pm 0.07$ & $1.41 \pm 0.02$ & $45 \pm 6~\%$\\
CONT2 & $1.71 \pm 0.02$ & $0.70 \pm 0.08$ & $1.85 \pm 0.03$ & $40 \pm 5~\%$\\
C\textsc{iv} & $5.01 \pm 0.06$ & $1.15 \pm 0.14$ & $5.13 \pm 0.06$ & $22 \pm 3~\%$\\
C\textsc{ii} & $2.66 \pm 0.04$ & $0.88 \pm 0.12$ & $2.79 \pm 0.04$ & $32 \pm 5~\%$\\
Si\textsc{iv}& $1.06 \pm 0.02$ & $0.45 \pm 0.06$ & $1.15 \pm 0.03$ & $39 \pm 4~\%$\\
O\textsc{iii}] & $1.73 \pm 0.03$ & $0.70 \pm 0.08$ & $1.86 \pm 0.04$ & $38 \pm 5~\%$\\
\hline
\end{tabular}
\end{table}

\clearpage

\begin{figure*}[htbp]
    \centering
    \includegraphics[width=\textwidth]{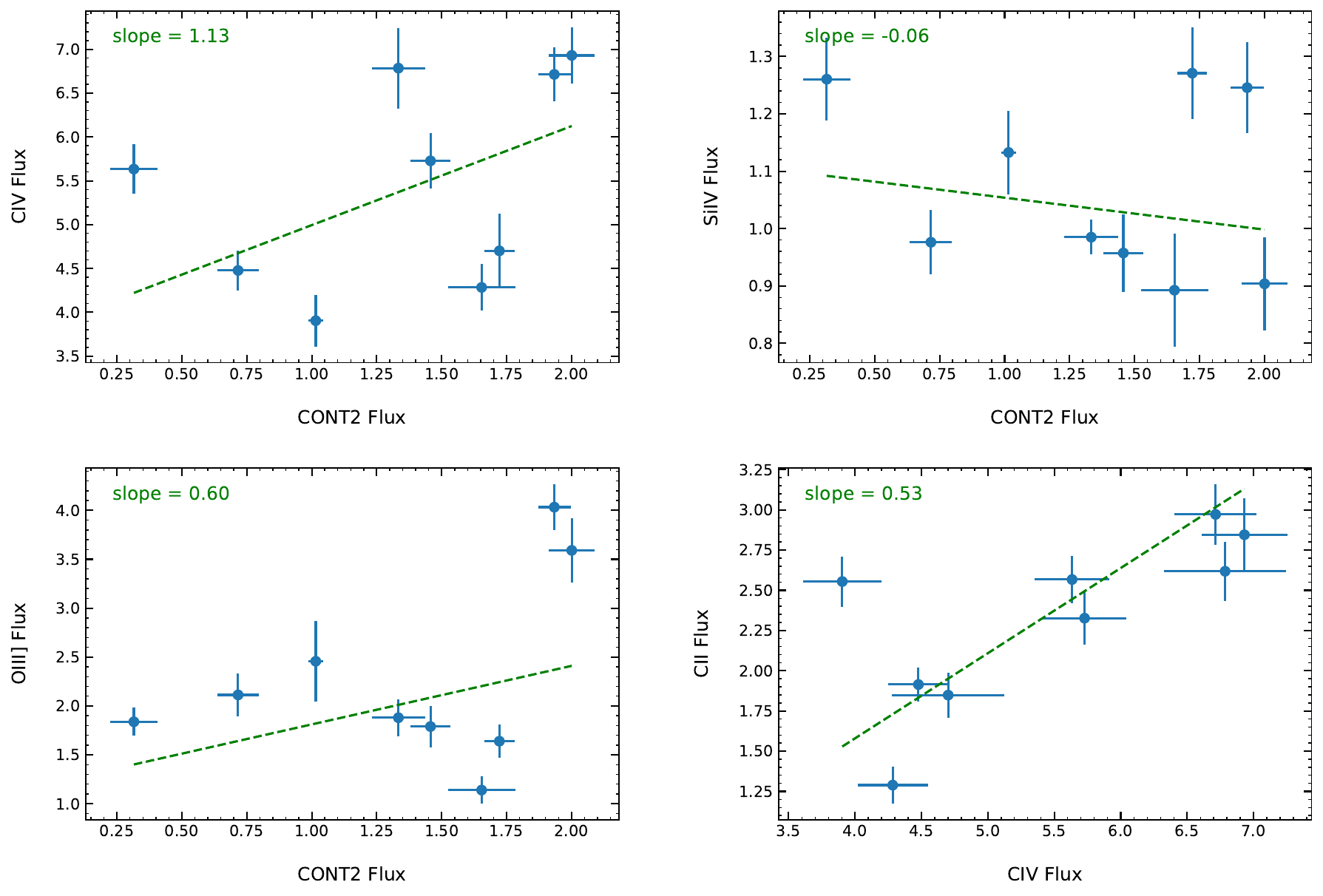}
    \caption{Correlation between UV continuum (CONT2: 1565-1633~\AA) and line fluxes ($10^{-13}\ \mathrm{erg\ s^{-1}\ cm^{-2}\ }$).}
    \label{Fig9}
\end{figure*}

\clearpage
\begin{table}[htbp]
\centering
\caption{C\textsc{iv}, C\textsc{ii}, Si\textsc{iv} and O\textsc{iii}] velocity widths due to doppler broadening for observations on 2-3 July, 1989. \label{tab5}}
\begin{tabular}{ccccc}
\hline
Observation & C\textsc{iv} Velocity & C\textsc{ii} Velocity & Si\textsc{iv} Velocity & O\textsc{iii}] Velocity\\
ID & Width (km/s) & Width (km/s) & Width (km/s) & Width (km/s)\\
\hline
SWP36598  &  $642.95 \pm 75.66$ & $593.26 \pm 113.71$ & $770.27 \pm 91.67$& $661.07 \pm 181.86 $\\
SWP36599  &  $794.43 \pm 78.87$ & $643.55 \pm 74.37$ & $587.28 \pm 157.84$& $986.04 \pm 250.37 $\\
SWP36600  &  $842.16 \pm 90.55$ & $857.87 \pm 99.09$ & $899.46 \pm 238.83$& $886.62 \pm 201.75 $\\
SWP36601  &  $933.40 \pm 98.40$ & $879.72 \pm 133.33$ & $820.15 \pm 339.35$& $1201.49 \pm 394.82 $\\
SWP36605  &  $963.10 \pm 99.53$ & $908.78 \pm 139.03$ & $253.80 \pm 107.79$& $447.23 \pm 165.73 $\\
SWP36606  &  $804.01 \pm 50.74$ & $682.09 \pm 64.59$ & $438.21 \pm 90.39$& $635.81 \pm 183.22 $\\
SWP36607  &  $1043.43\pm 126.53$& $796.83 \pm 123.93$ & $518.70 \pm 142.60$& $932.14 \pm 199.80 $\\
SWP36608  &  $680.67 \pm 117.76$& $688.09 \pm 119.28$ & $695.74 \pm 250.72$& $846.72 \pm 173.55 $\\
SWP36609  &  $797.71 \pm 67.96$ & $741.47 \pm 116.50$ & $545.78 \pm 178.21$& $1037.48 \pm 122.03 $\\
\hline
\end{tabular}
\end{table}
To investigate any potential correlation between the continuum flux and the line fluxes, we computed the Spearman rank, Pearson, and Kendall tau correlation coefficients, along with their associated p-values for various combinations of UV fluxes. These results are summarized in Table \ref{tab6}, with the corresponding correlation plot shown in Figure \ref{Fig9}. We haven't numerically correlated UV fluxes with optical magnitude as we do not have simultaneous observation data. For visualization, though we have extrapolated the optical light curve with a 6.67 hour period, it is not entirely accurate as the period is continuously decreasing in the order of $10^{-7}\ \mathrm{days\ year^{-1}}$  \citep{2018A&A...612A..91M}.
\begin{table}[htbp]
\centering
\caption{Correlation coefficients with their p-values for different combinations of continuum (CONT1: 1343-1390~\AA, CONT2: 1565-1633~\AA) and line fluxes. \label{tab6}}
\begin{tabular}{c|cc|cc|cc}
\hline
Flux Combination & Pearson r & p-value & Spearman rho & p-value & Kendall tau & p-value\\
\hline
CONT1 - C\textsc{iv} & 0.565 & 0.113 & 0.617 & 0.077 & 0.444 & 0.119 \\
CONT1 - Si\textsc{iv} & -0.454 & 0.220 & -0.583 & 0.099 & -0.444 & 0.119 \\
CONT1 - O\textsc{iii}] & 0.251 & 0.515 & 0.150 & 0.700 & 0.056 & 0.919 \\
CONT2 - C\textsc{iv} & 0.379 & 0.314 & 0.467 & 0.205 & 0.333 & 0.260 \\
CONT2 - Si\textsc{iv} & -0.214 & 0.580 & -0.217 & 0.576 & 0.222 & 0.477 \\
CONT2 - O\textsc{iii}] & 0.406 & 0.278 & 0.183 & 0.637 & 0.056 & 0.919 \\
C\textsc{iv} - C\textsc{ii} & 0.716 & 0.030 & 0.733 & 0.025 & 0.556 & 0.045 \\
\hline
\end{tabular}
\end{table}

\section{Discussion}
\label{sect:discussion}
UV line emission of contact binaries is generally attributed to its chromospheric behaviour. IUE Chromospheric lines have a temperature of $10^4$~K and transition region lines indicate $10^5$~K temperature from hot plasma \citep{1986ApJ...311..937V}. VW Cephei because of its rapid rotation and enhanced dynamo activity shows variation of flux in near and far UV. According to \citet{1980IAUS...88...39D}, the continuum emission is primarily due to heated photosphere and transition region activity. While the effective surface temperatures of the primary and secondary components are about 5050 K and 5342 K, respectively \citep{2018A&A...612A..91M}, which would typically produce a weak UV continuum, our spectral analysis shows that the UV continuum can be described by a power-law, which indicates that the origin of UV emission may include a combination of hot, optically thin plasma in the chromosphere and transition region, where temperatures exceed the photospheric values. 
In general, the C\textsc{iv} line at 1550~\AA, Si\textsc{iv} line at 1393~\AA~and the C\textsc{ii} line at 1336~\AA, are resonance lines formed under collisionally dominated conditions, originate in the highly active outer atmospheres. In contrast, O\textsc{iii}] line at 1660~\AA~is semi-forbidden transition, meaning this is a intercombination line, allowed only due to spin-orbit coupling. They occur in low-density, collisionally ionized regions, where the metastable upper states have time to decay radiatively. \citet{2014NewA...29...47S} attributed this variability of line fluxes to changes in local density and temperature, which are likely influenced by the presence and evolution of magnetically active regions, particularly on the primary star. Modeling studies using PHOEBE \citep{2005ApJ...628..426P} further support this interpretation. \citet{2018A&A...612A..91M} shows that the starspots on VW Cephei are located on the two opposite hemispheres of the primary star, aligned with the line connecting the two components. Hence there is a correlation between spottedness and the strength of the chromospheric activity. Figure \ref{Fig4} is consistent with the results of \citet{1986ApJ...311..937V}; at longer wavelengths (2580–3180 Å), the UV continuum follows the general trend of the optical light curve and \citet{2014NewA...29...47S}; fluxes for emission lines exhibit a clear correlation with the optical variations.

Table \ref{tab5} shows variability of velocity widths due to Doppler broadening reaching up to $\sim$1000 km/s. These values are considerably higher than the typical velocities expected from chromospheric turbulence or rotational broadening due to cool starspots. This discrepancy suggests that spot-related activity alone cannot account for the observed line broadening. Instead, the measured widths are more consistent with gas motion associated with Roche lobe overflow (RLOF). Using the system parameters of VW Cephei ($a=1.41\times10^6$ km and $M_1 = 1.13 M_{\odot}$ \citep{2018A&A...612A..91M}), the velocity of material flowing through the inner Lagrangian point (L1) can be approximated to $\sim$652 km/s, using a simple conservation of energy equation: $v = \sqrt{\frac{2GM_1}{r}}$, where $r=a/2$ is the approximate distance from the primary to L1. Thus the observed broadening matches well with the expected velocities from ballistic mass transfer, providing strong evidence that these UV line profiles are shaped by high-speed accretion flows rather than photospheric or chromospheric processes. Additionally, From Table \ref{tab6} we conclude that the variation of line fluxes are correlated with each other, but the dependence of line and continuum with each other is statistically insignificant. This supports the interpretation that the line-emitting plasma originates in dynamically heated accretion streams, distinct from the continuum originating photospheric or spot-dominated regions of the stars.

Using the mean flux values from Table~\ref{tab:SD}, we calculate the total ultraviolet (UV) flux to be 
\(
F_{\mathrm{UV}} = (13.45 \pm 0.08) \times 10^{-13} \, \mathrm{erg\,s^{-1}\,cm^{-2}}.
\)
Adopting a distance of \( d = 26.54 \, \mathrm{pc} \) \citep{2021A&A...649A...1G}, this corresponds to a UV luminosity of 
\(
L_{\mathrm{UV}} = 4 \pi d^2 F_{\mathrm{UV}} = (1.13 \pm 0.01) \times 10^{29} \, \mathrm{erg\,s^{-1}}.
\) For G-type and K-type stars, the typical bolometric correction factor in the UV is $\sim$ \(10^{-5}\). Applying this, we estimate the bolometric luminosity, \(
L_{\mathrm{Bol}} = (1.13 \pm 0.01) \times 10^{34} \, \mathrm{erg\,s^{-1}}
\). Using the standard formula for conservative mass transfer, \(\dot{M} = \frac{L_{\mathrm{Bol}} R_1}{G M_1}\) and adopting \( M_1 = 1.13 \, M_{\odot} \) and \( R_1 = 0.99 \, R_{\odot} \) \citep{2018A&A...612A..91M}, we derive a mass transfer rate of \(\dot{M} = (0.82 \pm 0.01) \times 10^{-7} \, M_{\odot}\,\mathrm{yr^{-1}}\), which agrees reasonably well with the value obtained from optical light curve studies, \(\dot{M}_{\mathrm{opt}} = (1.11 \pm 0.05) \times 10^{-7} \, M_{\odot}\,\mathrm{yr^{-1}},\) as reported by \citet{2018A&A...612A..91M}. Therefore, the large and variable line widths, rapid temporal modulation of UV fluxes, and weak correlation between line and continuum fluxes support the interpretation that the UV line profiles in VW Cephei are shaped by high velocity gas flows originating from Roche lobe overflow. The variability in line broadening likely reflects changes in the accretion stream geometry, impact conditions, or density fluctuations within the flow as the system evolves over the orbital cycle. However, to fully disentangle the physical origins of both the line and continuum emission, and to understand their variability future simultaneous multiwavelength observations will be essential.

\section{Conclusion}
\label{sect:conclusion}
A detailed spectroscopic investigation of the ultraviolet (UV) emission from the contact binary VW Cephei has been carried out using archival short-wavelength prime data from the International Ultraviolet Explorer (IUE). The analysis reveals prominent emission lines from ionized metal species, notably C\textsc{ii} and C\textsc{iv}, which are characteristic of late-type contact binaries. Emission lines due to O\textsc{iii}] and Si\textsc{iv} have also been reported for the first time for VW Cephei from our analysis. The UV flux variation appears to follow a similar trend to that observed in the optical light curves, suggesting a correlation between magnetic activity and UV variability. Rapid variability is observed in UV continuum (fractional rms of 40-45\%) as well as UV line fluxes (22-42\%) and they seem uncorrelated. This may imply that the line and continuum may have different geometric origin. However, simultaneous multi-wavelength observations are required for better understanding of the RL overflow geometry in VW Cephei. 
The change in velocity width of the line-emitting gas clouds is probably due to mass transfer between the binary components. To better understand the physical processes responsible for the emission features, especially the temporal behavior of the lines, additional multi-epoch UV observations are necessary.

\normalem
\begin{acknowledgements}
AB acknowledges DST, Government of India for the award of INSPIRE fellowship (IF230384) and the research facility of Department of Physics, IIT Hyderabad.
\end{acknowledgements}
  

\end{document}